\documentclass[pra,twocolumn,aps]{revtex4-1}
\usepackage{amsfonts,amssymb}
\begin{document}
\title{Entanglement and Complete Positivity: Relevance and
Manifestations in Classical Scalar Wave Optics}
\author{Arvind}
\email{arvind@iisermohali.ac.in}
\affiliation{Department of Physical Sciences,
Indian
Institute of Science Education \&
Research (IISER) Mohali, Sector 81 SAS Nagar,
Manauli PO 140306 Punjab India.}
\author{S. Chaturvedi}
\email{subhash@iiserbhopal.ac.in}
\affiliation{Department of Physics, 
Indian
Institute of Science Education \&
Research (IISER) Bhopal, Bhopal Bypass Road, Bhauri, Bhopal
462066 India}
\author{N. Mukunda}
\email{nmukunda@gmail.com}
\affiliation{
INSA C V Raman Research Professor,
Indian Academy of Sciences,
C V Raman Avenue, Sadashivanagar,
Bangalore 560080 India}
\begin{abstract}
Entanglement of states and Complete Positivity of maps are
concepts that have achieved physical importance with the
recent growth of quantum information science. They are
however mathematically relevant whenever tensor products of
complex linear (Hilbert) spaces are involved.  We present
such situations in classical scalar paraxial wave optics
where these concepts play a role: propagation
characteristics of coherent and partially coherent Gaussian
beams; and the definition and separability of the family of
Twisted Gaussian Schell Model (TGSM) beams.  In the former,
the evolution of the width of a projected one-dimensional
beam is shown to be a signature of entanglement in a
two-dimensional amplitude. In the latter, the partial
transpose operation is seen to explain key properties of
TGSM beams. 
\end{abstract}
\maketitle
\section{Introduction}
There have been some important and interesting developments
in quantum mechanics at the conceptual level which have
later been found to have parallels in classical wave optical
phenomena. Two well known instances are the discovery of the
geometric phase, and the existence of entanglement in
generic states of composite quantum systems.

The discovery of the quantum mechanical geometric phase in
1983--84 was in the context of adiabatic cyclic pure state
evolution governed by the time dependent Schr\"{o}dinger
equation~\cite{berry_1984}. However the earliest experimental
demonstration of this phase used a classical light wave
propagating along a coiled optical fibre, resulting in a
rotation of the plane of polarization~\cite{tomita_1986}. Somewhat later it
was realized that Pancharatnam's pioneering work in 1956
within classical polarization optics was an early example of
the (non adiabatic) geometric phase\cite{panch_1956}.

The concept of entanglement in states of composite quantum
systems goes back much earlier in time to the profound works
of Einstein, Podolsky and Rosen~\cite{epr_1935}, and of
Schr\"{o}dinger~\cite{schro_1936},
in 1935--1936\cite{neilsen_2000}. It has been realized more recently that
this property is relevant whenever one has to deal with
tensor products of (complex) linear vector spaces in any
physical situation, quantum or classical. When this occurs
in a classical context, it has been called nonquantum or
classical entanglement. An interesting recent study of
nonquantum entanglement involves the determination of all
physically realizable Mueller matrices in polarization
optics, which uses in an essential way the effects of
entanglement of polarization and spatial dependence of
classical electromagnetic wave fields~\cite{simon_2010}. Several other
more recent studies have analysed the concept of classical
entanglement further~\cite{xiao_2011}.

The purpose of the present work is to bring out the
importance of classical entanglement in chosen scalar wave
optical situations, thus not involving polarization at all.
It turns out that certain features of these situations are
understood much better in terms of their classical
entanglement properties.  The first situation that we
consider involves the propagation characteristics of
coherent and partially coherent paraxial Gaussian wave
fields. Here it is shown that the behaviour of a
one-dimensional beam width during propagation acts as an
entanglement witness in a two-dimensional amplitude,
understood appropriately. The second case is the study of a
class of partially coherent (paraxial) beams known as
Twisted Gaussian Schell Model (TGSM) beams. We show that by
using the action of the partial transpose operation  -- a
positive but not completely positive map used as
entanglement witness -- on these beams, their physical
properties -- range of parameters defining them,
separability as incoherent mixtures of product beams -- are
understood much better than previously.  
While the connection of entanglement with non-locality is
missing  in these classical situations, the notion of
entanglement provides insights into the structure of such beams
and their propagation.
Section~\ref{gaussian_beams} deals with the paraxial
Gaussian wave fields and beams, Section~\ref{TGSM} with the
TGSM family.  Section~\ref{conc} contains concluding
remarks.
\section{Gaussian beam propagation -- width behaviour as
entanglement witness}
\label{gaussian_beams}

The scalar paraxial wave equations in one and two
dimensions, with wavelength $\lambda
(\lambdabar=\lambda/2\pi)$ and propagating along the
$z$-axis, are

\begin{eqnarray}
i\frac{\partial}{\partial z}\psi(x;
z)&=&-\frac{\lambdabar}{2}\frac{\partial^2}{\partial
x^2}\psi(x; z),
\label{wave-1d}\\
i\frac{\partial}{\partial z}\Psi(x, y;
z)&=&-\frac{\lambdabar}{2}\left(\frac{\partial^2}{\partial
x^2}+\frac{\partial^2}{\partial y^2}\right)\Psi(x, y;
z).
\label{wave-2d}
\end{eqnarray}

We begin with the one-dimensional case.
\subsection*{Coherent and partially coherent one-dimensional
Gaussian beams~\cite{seigman_1986}}
We start with a real centered coherent Gaussian amplitude in
the waist plane $z=0$:
\begin{equation}
\psi(x; 0)=\left(\frac{2I}{\pi w^2}\right)^{1/4}{\rm e}^{-x^2/w^2},
\label{gaussian_beam}
\end{equation}
$w$ = width, $I$ = intensity measure.
Solving Eqn.~(\ref{wave-1d}) leads to
\begin{eqnarray}
\psi(x; z)&=&\left(\frac{2I}{\pi w^2}\right)^{1/4}
\left(1+\left(\frac{2\lambdabar
z}{w^2}\right)^2\right)^{-1/4}
\nonumber \\
&&\quad\quad \times
 \exp\left({\frac{-i}{2}
\tan^{-1}\left(\frac{2\lambdabar z}{w^2}\right)}\right)
\nonumber\\
&& \quad\quad \times \exp\left(-x^2/w^2\left(1+\frac{2i\lambdabar
z}{w^2}\right)\right).
\end{eqnarray}
(The $x$-independent exponent is the Guoy phase). This expression becomes more compact in terms of the Rayleigh range $z_R$, evolving width $w(z)$ and radius of curvature $R(z)$:
\begin{eqnarray}
z_R&=&w^2/2\lambdabar;\nonumber\\
w(z)&=&w\left(1+\left(\frac{z}{z_R}\right)^2\right)^{1/2};\nonumber\\
R(z)&=&-\left(z+\frac{z_R^2}{z}\right).
\label{beam_parameters}
\end{eqnarray}
Then
\begin{eqnarray}
\psi(x;z)&=&\left(\frac{2I}{\pi w(z)^2}\right)^{1/4}
 \exp\left({\frac{-i}{2}
\tan^{-1}\left(\frac{z}{z_R}\right)}\right)\nonumber\\
&& \quad\quad\times \exp\left(\frac{-x^2}{w(z)^2}- \frac{ix^2}{2\lambdabar R(z)}\right).
\end{eqnarray}
In the waist plane $|R(0)|=\infty$, and $z_R$ is the distance over which the width increases by a factor $\sqrt{2}$:
\begin {equation}
w(z_R)=\sqrt{2} w.
\end{equation}
In the coherent case, $z_R$ \emph{depends on $w$ alone}.

To introduce partial coherence we bring in the two-point
correlation function $\Gamma(x, x'; z)$. The so-called
Gaussian Schell Model (GSM) beam begins in the waist plane
as~\cite{simon_1984}
\begin{eqnarray}
\Gamma(x, x'; 0)=\left(\frac{2I}{\pi w^2}\right)^{1/2}\exp
\left(-\frac{x^2+x'^2}{w^2}-\frac{(x-x')^2}{2\delta^2}\right).
\nonumber \\
\label{twopt_partial0}
\end{eqnarray}

As in Eqn.~(\ref{gaussian_beam}), $w$ is the beam width; and
$\delta$ is the coherence length. In the coherent case
$\delta=\infty$. To find $\Gamma(x, x'; z)$ we use
Eqn.~(\ref{wave-1d})
with respect to $x$, and its complex conjugate with
respect to $x'$. Now the Rayleigh range turns out to depend
on \emph{both $w$ and} $\delta$:
\begin{equation}
z_R={\frac{w^2}{2\lambdabar}}\left({1+\frac{w^2}{\delta^2}}\right)^{-\frac{1}{2}}.
\end{equation}
This is smaller than the coherent case value in
Eqn.~(\ref{beam_parameters}); and the two-point function turns out to be
\begin{eqnarray}
\Gamma(x, x'; z)&=&\left(\frac{2I}{\pi
w(z)^2}\right)^{1/2}\nonumber \\
&&\!\!\!\!\!\!\!\!
\times \exp \left(-\frac{x^2+x'^2}{w(z)^2}-\frac{(x-x')^2}{2\delta(z)^2}-i\frac{(x^2-x'^2)}{2\lambdabar R(z)}\right),\nonumber\\
w(z)&=&w\left(1+(\frac{z}{z_R})^2\right)^{1/2}
\nonumber \\
&=& w\left(1+(\frac{2\lambdabar z}{w^2})^2(1+\frac{w^2}{\delta^2}) \right)^{1/2},\nonumber\\
\delta(z)&=&\delta\left(1+(\frac{z}{z_R})^2\right)^{1/2},\nonumber\\
R(z)&=&-(z+\frac{z_R^2}{z}).
\label{twopt_partialz}
\end{eqnarray}
Now over the Rayleigh range both width and coherence length
increase by a factor $\sqrt{2}$:
\begin{equation}
w(z_R)=\sqrt{2}w,\quad \delta(z_R)=\sqrt{2}\delta.
\end{equation}
Comparing with the coherent case, since now $z_R$ has
decreased, it is seen that the finite coherence length
$\delta$ results in an increased divergence of the beam:
\begin{eqnarray}
{\rm partial~coherence} &\Rightarrow& {\rm finite~} \delta
\nonumber \\
&\Rightarrow& {\rm decreased~} z_R
\nonumber \\
&\Rightarrow& {\rm increased~ beam~ divergence.}
\end{eqnarray}
This result will now be shown to serve as an entanglement witness in a two-dimensional coherent Gaussian beam.

\subsection*{The two-dimensional case}
For a coherent two-dimensional Gaussian amplitude $\Psi(x,
y; z)$ we consider the product of two one-dimensional
centered Gaussian amplitudes, in $x$ and in $y$, each
obeying Eqn.~(\ref{wave-1d}); this then obeys
Eqn.~(\ref{wave-2d}). We assume a
common waist plane at $z=0$, \emph{but with unequal widths}:
\begin{eqnarray}
\Psi(x, y; 0)&=&\left(\frac{2I_1}{\pi w_1^2}\cdot
\frac{2I_2}{\pi w_2^2}\right)^{1/4}\cdot \exp
\left(\frac{-x^2}{w_1^2}-\frac{y^2}{w_2^2}\right),
\nonumber \\
&&w_1>w_2.
\end{eqnarray}
The elliptic waist has principal axes $x$ and $y$.
Solving Eqn.~(\ref{wave-2d}) we have for general 
$z$ by construction the
product form
\begin{eqnarray}
&&\Psi(x, y; z)=\left(\frac{2I_1}{\pi w_1(z)^2}
\frac{2I_2}{\pi w_2(z)^2}\right)^{1/4} \nonumber \\
&&
\times
\exp\left({\frac{-i}{2}\left(\tan^{-1}
\left(\frac{z}{z_{1R}}\right)+
\tan^{-1}\left(\frac{z}{z_{2R}}
\right)\right)}\right)\nonumber\\
&& \times\exp
\left(\frac{-x^2}{w_1(z)^2}-\frac{y^2}{w_2(z)^2}-
\frac{i}{2\lambdabar}\left(\frac{x^2}{R_1(z)}+\frac{y^2}{R_2(z)}
\right)\right);\nonumber\\
&&z_{aR}=w_a^2/2\lambdabar,\nonumber\\
&&w_a(z)=w_a\left(1+\left(\frac{z}{z_{aR}}\right)^2\right)^{1/2}=
w_a\left(1+\left(\frac{2\lambdabar
z}{w_{a}^2}\right)^2\right)^{1/2},\nonumber\\
&&R_a(z)=-\left(z+\frac{z_{aR}^2}{z}\right),\quad a=1, 2.
\end{eqnarray}
This is coherent, centered, Gaussian and elliptic, with
principal axes $x$ and $y$, for all $z$. Since
$z_{1R}>z_{2R}, w_1(z)$ expands at a slower rate than
$w_2(z)$.

Now make an anti-clockwise rotation by angle $\theta$ in the
$x$--$y$ plane to $x'$--$y'$ axes:
\begin{equation}
\left(\begin{array}{c}x'\\ y'\end{array}\right)
=\left(\begin{array}{cc}\cos \theta & \sin\theta\\
-\sin\theta &\cos\theta\end{array}\right)
\left(\begin{array}{c}x\\ y\end{array}\right).
\end{equation}
On account of the assumed anisotropy $w_1>w_2$, the two-dimensional amplitude expressed in the new variables,
\begin{equation}
\Psi'(x', y';z)=\Psi(x,y;z),
\label{rotated_field}
\end{equation}
is no longer of the product form in $x'$ and $y'$, i.e.
\emph{it is definitely entangled} as a function of these
variables. As we now show, this entanglement can be revealed
by a family of one-dimensional measurements.

Suppose we limit all measurements to
quantities--intensities, widths, $\ldots$ -- in the $x'$
direction, and their $z$-dependences. To `trace over'
dependences on $y'$, we first construct the product
two-point function in the new variables
\begin{equation}
\Gamma'(x', y'; x'', y''; z)=\Psi'(x', y'; z) \Psi'(x'',
y''; z)^*,
\end{equation}
then set $y''=y'$ and integrate over $y'$ to obtain the
one-dimensional two-point correlation function
\begin{equation}
\Gamma''(x',  x''; z)=
\int_{-\infty}^{\infty}{\rm d}y'\ \Gamma'(x', y'; x'', y'; z).
\label{twopt_reduced}
\end{equation}
It is easy to see that this is necessarily a centered
one-dimensional GSM beam as in
Eqn.(\ref{twopt_partial0},\ref{twopt_partialz}). The width
$w'(z)$ is:
\begin{eqnarray}
&& w'(z)^2=4\langle x'^2\rangle
=\frac{\displaystyle 4 \int_{-\infty}^{\infty}{\rm d}x'x'^2
\Gamma''(x', x'; z)}{\displaystyle \int_{-\infty}^{\infty}{\rm d}x' 
\Gamma''(x', x'; z)} \nonumber\\
&& =\frac{\displaystyle 4\int \int {\rm d}x{\rm d}y 
(x\cos \theta+y\sin \theta)^2|\Psi(x, y; z)|^2}{
\displaystyle \int \int {\rm d}x{\rm d}y|\Psi(x, y; z)|^2}\nonumber\\
&& =\cos^2\theta\ w_1(z)^2+\sin^2 \theta\ w_2(z)^2\nonumber\\
&&
=w_1^2\left(1+\left(\frac{z}{z_{1R}}\right)^2\right)\cos^2\theta+w_2^2
\left(1+\left(\frac{z}{z_{2R}}\right)^2\right)\sin^2\theta.
\nonumber \\
\end{eqnarray}
At $z=0$ this is
\begin{equation}
w'(0)^2=w_1^2\cos^2\theta+w_2^2\sin^2\theta,
\end{equation}
so to compare with the one-dimensional GSM result
in~(\ref{twopt_partialz}) we examine the ratio
\begin{eqnarray}
&&\left(\frac{w'(z)}{w'(0)}\right)^2=
1+\left(\left(\frac{w_1}{z_{1R}}\right)^2\cos^2\theta
+\left(\frac{w_2}{z_{2R}}\right)^2\sin^2\theta\right)
\frac{z^2}{w'(0)^2}\nonumber\\
&&=1+w'(0)^2\left(\frac{\cos^2\theta}{w_1^2}+
\frac{\sin^2\theta}{w_2^2}\right)
\left(\frac{2\lambdabar z}{w'(0)^2}\right)^2.
\end{eqnarray}
On account of
\begin{eqnarray}
w'(0)^2\left(\frac{\cos^2\theta}{w_1^2}+
\frac{\sin^2\theta}{w_2^2}\right) =
1+\left(\frac{w_1^2-w_2^2}{2w_1w_2}\right)^2
\sin^2 2\theta, \nonumber \\
\end{eqnarray}
the above ratio simplifies to
\begin{eqnarray}
&& \left(\frac{w'(z)}{w'(0)}\right)^2=1+ \left(\frac{2\lambdabar z}{w'(0)^2}\right)^2\left(1+\frac{w'(0)^2}{\delta^2}\right),\nonumber\\
&& \frac{w'(0)}{\delta}=\frac{w_1^2-w_2^2}{2w_1w_2}|\sin 2\theta|.
\end{eqnarray}
This is to be compared with the one-dimensional GSM result
in Eqn.~(\ref{twopt_partialz}). 
Since $w_1>w_2$ (and assuming $\theta\ne 0,
\pi/2$) the one-dimensional GSM beam $\Gamma''(x', x''; z)$
in Eqn.~(\ref{twopt_reduced}) has finite coherence length $\delta$. In turn
this means that in the two-dimensional coherent Gaussian
amplitude $\Psi'(x', y'; z)$ of Eqn.~(\ref{rotated_field}) there is
entanglement of $x'$ and $y'$, and \emph{this is directly
seen in the propagation behaviour of the beam width $w'(z)$
in the direction $x'$ in the transverse plane}.
Equivalently, entanglement between $x'$ and $y'$ in
$\Psi'(x', y'; z)$ leads to partial coherence in the reduced
one-dimensional $x'$-mode which is of GSM type, and this can
be measured through the propagation behaviour of beam width
$w'(z)$.
This situation is analogous to bipartite pure entangled
quantum states, where the reduced density operator of any of the
components is a mixed state and this mixedness is used as an
entanglement witness for the original combined state.
\section{Gaussian Schell Model beams -- 
anisotropy, twist and partial transpose operation}
\label{TGSM}
Now we look at the application of a positive but not
completely positive (CP) map -- the partial transpose (PT)
operation\cite{neilsen_2000} -- on a well-studied family of two-dimensional
optical two-point functions. For a one-dimensional partially
coherent beam the defining properties of the two-point
function $\Gamma(x; x')$ are hermiticity and positive
semidefiniteness:
\begin{eqnarray}
&& \Gamma(x; x')^*= \Gamma(x'; x),\nonumber\\
&& \{\Gamma(x; x')\}\ge 0\ \hbox{as a one-dimensional integral kernel}.
\end{eqnarray}
These are obviously preserved by the transposition map
\begin{eqnarray}
&& \Omega_x: \Gamma(x; x')\rightarrow 
\tilde{\Gamma}(x; x')= \Gamma(x'; x), \nonumber\\
&& \Gamma\ {\rm physical}
\Leftrightarrow \tilde{\Gamma} 
\ {\rm physical}.
\end{eqnarray}
However when $\Omega_x$ is trivially extended to the PT
operation $\Omega_x\times \mathbb{I}_y$ on two-dimensional
two-point functions, 
\begin{eqnarray} \Omega_x\times
\mathbb{I}_y: \Gamma(x, y; x', y')\rightarrow
\tilde{\Gamma}(x, y; x', y')= \Gamma(x', y; x, y'),
\nonumber \\
\label{pt_twopt}
\end{eqnarray}
while hermiticity is preserved the positive
semi-definiteness may be lost. As two dimensional integral
kernels,
\begin{equation}
\{\Gamma(x, y; x', y')\}\ge 0\not\Rightarrow 
\{\tilde{\Gamma}(x, y; x', y')\}\ge 0.
\end{equation}
Thus $\Omega_x$ is not CP. Furthermore we have the consequence
\begin{equation}
\{\tilde{\Gamma}(x, y; x', y')\}\not\geq 0
\Rightarrow \{\Gamma(x, y; x', y')\}\ {\rm entangled}.
\end{equation}
This is mathematically similar to the case of bipartite
density operators in quantum mechanics, where entangled
states can turn into `non-states' under the action of PT
operation.
\subsection*{The AGSM family of beams} We examine this in
the case of a family of two-dimensional centered Gaussian
Schell Model beams with interesting
properties~\cite{simon_1985}. Denote two general points $(x,
y)^T, (x', y')^T$ in the transverse plane by $\bm{\rho},
\bm{\rho}'$.  The most general Anisotropic Gaussian Schell
Model (AGSM) beam is described by the two-point function
\begin{eqnarray}
&& \Gamma^{(AGSM)}(\bm{\rho}; \bm{\rho}')=\nonumber  \\
&&\frac{I}{2\pi}
(\det L)^{1/2}\exp\left\{-\frac{1}{4}\bm{\rho}^T\ L \ 
\bm{\rho}-\frac{1}{4}\bm{\rho}'^T\ L \bm{\rho}' \right.\nonumber\\
&&\left.  -\frac{1}{2}(\bm{\rho}-\bm{\rho}')^T 
M(\bm{\rho}-\bm{\rho}')-
\frac{i}{2\lambdabar}
(\bm{\rho}-\bm{\rho}')^TK(\bm{\rho}+\bm{\rho}') \right\}.
\label{agsm_beam}
\end{eqnarray}
Here $L, M, K$ are real $2\times 2$ matrices obeying
\begin{equation}
L^T=L>0,\quad M^T=M\ge 0,
\label{lm_conditions}
\end{equation}

and another condition developed below. Apart from the total
intensity $I$, this is a ten-parameter family: $L$
determines the intensity distribution, $M$ the partial
coherence, and $K$ the phase. The exponent on the right
in~(\ref{agsm_beam}) is the most general hermitian quadratic expression
in $\bm{\rho}$ and $\bm{\rho}'$.

It is convenient to pass from the physical space 
description~(\ref{agsm_beam})
to an equivalent Wigner distribution description on
`optical phase space'. Points in this space correspond to
real four component column vectors
\begin{equation}
\xi=\left(\begin{array}{c}\bm{\rho}\\ {\bf
p}\end{array}\right), \quad {\bf
p}=\left(\begin{array}{c}p_x\\ p_y\end{array}\right),
\end{equation}
with ${\bf p}$ dimensionless. 
The Wigner distribution $W(\xi)$ corresponding to $\Gamma^{(AGSM)}(\bm{\rho}; \bm{\rho}')$ is
\begin{eqnarray}
&&W(\xi)=
(2\pi\lambdabar)^{-2}
\nonumber \\
&&\times
\int {\rm d}^2\bm{\rho}' 
{\rm e}^{\frac{\displaystyle -i{\bf p}\cdot \bm{\rho}'
}{\displaystyle \lambdabar}} \,\Gamma^{(AGSM)}{({\bm{\rho}}+
\frac{1}{2}\bm{\rho}'; \bm{\rho}-
\frac{1}{2}\bm{\rho}')}\nonumber\\
& =&\frac{I}{(2\pi)^2}
(\det V)^{-1/2}\cdot 
\exp\left(-\frac{1}{2}\xi^T V^{-1}\xi\right),\nonumber\\
V&=&\left(\begin{array}{cc}L^{-1}& -L^{-1}K^T\\
-KL^{-1}& \quad KL^{-1}K^T+\lambdabar^2(\frac{1}{4} L+M)\end{array}\right).
\label{wigner}
\end{eqnarray}
The real symmetric positive definite $4\times 4$ matrix $V$
describes this (centered) Gaussian phase space distribution,
hence the original beam~(\ref{agsm_beam}), completely. It is the matrix
of second moments or variances of the beam:
\begin{equation}
V_{ab}=\frac{1}{I}\int {\rm d}^4\xi\ 
\xi_a\xi_b\ W(\xi),\quad a,b=1, 2, 3, 4.
\end{equation}
The condition that $\Gamma^{(AGSM)}(\bm{\rho};
\bm{\rho}')$ be a positive semidefinite integral kernel is
expressed by the `optical uncertainty
principle'~\cite{simon_1987}
\begin{equation}
\qquad V+\frac{i}{2}\lambdabar \beta\ge 0;\quad
\beta=\left(\begin{array}{cc}0& 
\mathbb{I}_{2\times 2}\\ 
-\mathbb{I}_{2\times 2}& 
0\end{array}\right).
\label{vmatrix_conditions}
\end{equation}
Thus the complete set of conditions for physical
acceptability are
Eqn.~(\ref{lm_conditions},\ref{vmatrix_conditions}).
\subsection*{The TGSM subfamily}
Now we limit ourselves to two subfamilies in the
ten-parameter AGSM family. The first is the four-parameter
Twisted Gaussian Schell Model (TGSM)
subfamily~\cite{simon_1993}:

\begin{eqnarray}
&&L=\frac{4}{w^2}\cdot \mathbb{I}, \,\, M=\frac{1}{\delta^2}\cdot 
\mathbb{I},\,\, K=\frac{1}{R}\cdot \mathbb{I}+iu\sigma_2:\nonumber\\
&& \Gamma^{(TGSM)}(\bm{\rho}; \bm{\rho}')=
\frac{2I}{\pi w^2}\cdot 
\exp\left\{-\frac{1}{w^2}(\bm{\rho}^2+\bm{\rho}'^2)\right.
\nonumber \\
&&- \left. \frac{1}{2\delta^2}(\bm{\rho}-\bm{\rho}')^2-
\frac{i}{2\lambdabar R}(\bm{\rho}^2-\bm{\rho}'^2)
 -\frac{iu}{\lambdabar}(xy'-yx')\right\}.
\label{tgsm_twopt}
\end{eqnarray}
This is the most general AGSM beam invariant under proper
SO(2) rotations about the propagation direction. Apart from
total intensity $I$, the four independent parameters are
intensity width $w$, coherence length $\delta$, radius of
curvature $R$ and twist $u$.

With these choices for $L, M, K$ the variance matrix $V$ of
Eqn.~(\ref{wigner}) is
\begin{eqnarray}
&& V=\left(\begin{array}{cccc}a&0&b & c\\ 0 & a& -c & b\\ b& -c & d& 0\\ c & b &0 & d\end{array}\right),\nonumber\\
&& a=\frac{w^2}{4}, b=\frac{-w^2}{4R},
c=\frac{uw^2}{4}, \nonumber \\
&&d=\lambdabar^2\left(\frac{1}{w^2}+\frac{1}{\delta^2}\right)+
\frac{w^2}{4}\left(u^2+\frac{1}{R^2}\right).
\label{v_abcd}
\end{eqnarray}
The hermitian matrix $V+\frac{i}{2}\lambdabar \beta$ has
real eigenvalues; we must ensure these are all nonnegative.
The characteristic equation for the eigenvalues $\mu$ is
\begin{eqnarray}
\det\left(V+\frac{i}{2}\lambdabar \beta -\mu\cdot
\mathbb{I}_{4\times 4}\right)
=((a-\mu)(d-\mu))^2
\nonumber \\
-2(a-\mu)(d-\mu)(b^2+c^2+\frac{\lambdabar^2}{4})
\nonumber\\
 +((b+\frac{i}{2}\lambdabar)^2+c^2) ((b-\frac{i}{2}\lambdabar)^2+c^2)=0,
\end{eqnarray}
from which the eigenvalues $\mu$ are easily determined:
\begin{eqnarray}
&& (a-\mu)(d-\mu)=b^2+c^2+\frac{\lambdabar^2}{4}+\varepsilon\lambdabar c,\quad \varepsilon=\pm 1;\nonumber\\
&& 2\mu=a+d+\varepsilon'\{(a-d)^2+4(b^2+c^2+\frac{\lambdabar^2}{4}+\varepsilon \lambdabar c)\}^{1/2},\nonumber\\
&& \varepsilon=\pm 1, \varepsilon'=\pm 1.
\label{eigen_tgsm}
\end{eqnarray}
The condition that they be all nonnegative is
\begin{eqnarray}
(a+d)^2&\ge& (a-d)^2+4(b^2+c^2+\frac{\lambdabar^2}{4}+\varepsilon \lambdabar c),\nonumber\\
{\rm i.e.}\quad ad&\ge& b^2+c^2+\frac{\lambdabar^2}{4}+\varepsilon \lambdabar c,\quad \varepsilon=\pm 1;\nonumber\\
{\rm i.e.}\quad |u|&\le& \lambdabar/\delta^2.
\end{eqnarray}
Therefore the two-point function~(\ref{tgsm_twopt})
is a physically realizable TGSM beam only when
\begin{equation}
w>0, \quad \delta\ge 0, \quad |u|\le \lambdabar/\delta^2.
\end{equation}
\subsection*{The $\Gamma^{{\rm (curv)}}$ subfamily}
The second four-parameter subfamily is chosen so that,
formally, it is related to the TGSM subfamily by the PT map
$\Omega_x\times\mathbb{I}_y$ of Eqn.~(\ref{pt_twopt}). We denote it by
$\Gamma^{{\rm (curv)}}$, with the choices of $L, M, K$ being
\begin{equation}
L=\frac{4}{w^2}\cdot \mathbb{I}, M=\frac{1}{\delta^2}\cdot \mathbb{I}, K=-u\sigma_1-\frac{1}{R}\sigma_3.
\label{lm_tgsm}
\end{equation}
Compared to the previous choices~(\ref{tgsm_twopt}), only $K$ is
different, being now symmetric traceless. The corresponding
two-point functions are:
\begin{eqnarray}
&& \Gamma^{{\rm (curv)}}(\bm{\rho}; \bm{\rho}')=\frac{2I}{\pi
w^2}\cdot
\exp\{-\frac{1}{w^2}(\bm{\rho}^2+\bm{\rho}'^2)-\frac{1}{2\delta^2}(\bm{\rho}-\bm{\rho}')^2\nonumber\\
&& +\frac{i}{2\lambdabar
R}(x^2-y^2-x'^2+y'^2)+\frac{iu}{\lambdabar}(xy-x'y')\}.
\label{tcurr}
\end{eqnarray}
We must now check for any acceptability conditions on $w,
\delta, R$ and $u$. From Eqn.~(\ref{wigner},\ref{lm_tgsm}) the variance
matrix, written $V'$, is
\begin{eqnarray}
&& V'= \left(\begin{array}{cccc}a'&0&b' & c'\\ 0 & a'& c' & -b'\\ b'& c' & d'& 0\\ c' & -b'& 0 & d'\end{array}\right),\nonumber\\
&& a'=\frac{w^2}{4}, b'=\frac{w^2}{4R},
c'=\frac{uw^2}{4},\nonumber\\ &&
d'=\lambdabar^2\left(\frac{1}{w^2}+\frac{1}{\delta^2}\right)+
\frac{w^2}{4}\left(u^2+\frac{1}{R^2}\right).\end{eqnarray}
Compared to Eqn.~(\ref{v_abcd}) for $V$, there are some
sign differences; and while $a'=a, c'=c, d'=d$, we have
$b'=-b$. The characteristic equation for the eigenvalues
$\mu'$ of $V'+\frac{i}{2}\lambdabar \beta$ is:
\begin{eqnarray}
&&\det \left(V'+\frac{i}{2}\lambdabar \beta-\mu'\cdot
\mathbb{I}_{4\times 4}\right)=
\nonumber \\
&&\quad\quad((a'-\mu')(d'-\mu')-b'^2-c'^2-\frac{\lambdabar^2}{4})^2=0,
\end{eqnarray}
so the eigenvalues are doubly degenerate given by
\begin{eqnarray}
2\mu'&=&a'+d'+\varepsilon'\left\{(a'-d')^2+
4(b'^2+c'^2+\frac{\lambdabar^2}{4})\right\}^{1/2},\nonumber
\\
\varepsilon'&=&\pm 1.
\end{eqnarray}
We now find that, in contrast to Eqn.~(\ref{eigen_tgsm}), all these
eigenvalues are non negative, in particular with no
inequalities on $|u|$. Thus for all $w>0, \delta\ge 0$, real
$R$ and real $u$, $\Gamma^{{\rm (curv)}}(\bm{\rho}; \bm{\rho}')$
of Eqn.~(\ref{tcurr}) is physical.
\subsection*{The effect of the PT operation}
We now put together these two sets of results. As just seen,
the beam $\Gamma^{{\rm (curv)}}(\bm{\rho}; \bm{\rho}')$ of
Eqn.~(\ref{lm_tgsm}) 
is physical for all real $R$ and $u$ ($w>0, \delta\ge
0$ always). Upon applying the PT operation to it, we
formally obtain the right hand side of the TGSM
beam~(\ref{tgsm_twopt});
this is the basis for the choices of $L, M, K$ in Eqn.~(\ref{lm_tgsm}). 
However we know that for the TGSM beam to be
physical, the condition $|u|\le \lambdabar /\delta^2$ must
be obeyed. Therefore we arrive at:

\begin{eqnarray}
|u|\le \lambdabar /\delta^2: 
&&(\Omega_x\times \mathbb{I}_y) \Gamma^{{\rm
(curv)}}(\bm{\rho}; 
\bm{\rho}')=\Gamma^{(TGSM)}(\bm{\rho}; \bm{\rho}'),\nonumber\\
&&(\Omega_x\times \mathbb{I}_y) \Gamma^{(TGSM)}(\bm{\rho};
\bm{\rho}')=\Gamma^{{\rm (curv)}}(\bm{\rho}; \bm{\rho}');\nonumber\\
|u|> \lambdabar /\delta^2: &&(\Omega_x\times \mathbb{I}_y) 
\Gamma^{{\rm (curv)}}(\bm{\rho}; \bm{\rho}')=\ 
{\rm unphysical}. 
\end{eqnarray}
This can be combined with the well-known result for Gaussian
states of continuous variable systems\cite{simon_2000}:
positivity/negativity under PT implies
separability/entanglement. Therefore:

\begin{eqnarray}
|u|\le \lambdabar /\delta^2: \Gamma^{(TGSM)}, 
\Gamma^{{\rm (curv)}}\ {\rm both\ separable};\nonumber\\
|u|> \lambdabar /\delta^2: \Gamma^{{\rm (curv)}}\, 
{\rm non\ separable}.
\end{eqnarray}
The separability of TGSM beams -- a `superposition model' --
was shown soon after these beams were
discovered~\cite{gori_1994}. This
property is now understood in a new and deeper manner. The
nonexistence of TGSM beams for $|u|> \lambdabar /\delta^2$
gets related to nonseparability or entanglement of
$\Gamma^{{\rm (curv)}}$ in this parameter range. We also
appreciate the physical differences between the phases
appearing in the two cases, 
Eqn.~(\ref{tgsm_twopt},\ref{tcurr}): while the
latter can be created by suitable concavo--convex lenses,
the twist phase is more subtle and cannot be created in such
a simple manner.

\section{Concluding remarks}
\label{conc}
We have presented two examples from classical scalar
paraxial wave optics which bring out the relevance of
entanglement and of the non CP partial transpose operation.
Although traditionally these notions are viewed as quantum
mechanical in origin they are present in these classical contexts.
In the terminology explained in the last
of references~\cite{xiao_2011}, these are instances of intrasystem
entanglement. In the first case, the presence of
entanglement in a coherent anisotropic Gaussian amplitude is
shown to result in partial coherence in a projected
one-dimensional beam; this can then be detected by
one-dimensional measurements of the width in the projected
beam. The second example involves the well studied family of
TGSM beams. These beams are physically realisable only when
the twist parameter $u$ is bounded above in magnitude by an
expression inversely proportional to the square of the
coherence length. In particular, no twist phase can be
impressed upon a fully coherent beam. In contrast, the beams
$\Gamma^{{\rm (curv)}}$ are physical for all values of $u$; but
they change from being separable to being entangled as $|u|$
passes through $\lambdabar/\delta^2$. 
This is reminiscent of the behaviour of the Gaussian states
of the two mode electromagnetic fields and the so-called
Werner states much studied in quantum information theory.

The role of the bound on the
twist phase, and the fact that TGSM beams are always
separable, are explained in a new light using the action of
the partial transpose operation on these beams, carrying
them to the beams $\Gamma^{{\rm (curv)}}$.
We have shown that entanglement in scalar optical fields at the
classical level exists and its analysis provides useful
insights. In our view the situation needs to be investigated
further to find possible applications of this phenomenon of classical
entanglement. 
\section*{Acknowledgements}
Arvind acknowledges funding from DST India under Grant No.
EMR/2014/000297.  NM thanks the Indian National Science
Academy for enabling this work through the INSA C V Raman
Research Professorship.  NM thanks Professor R. Simon for
enlightening discussions on the topics dealt with in this
paper.



\begin{thebibliography}{99}
\bibitem{berry_1984} M. V. Berry, `Quantal Phase Factors
Accompanying Adiabatic Changes', Proc. Roy Soc. (London)
\textbf{A392}, 45--57 (1984).
\bibitem{tomita_1986} A. Tomita and R. Chiao, `Observation
of Berry's Topological Phase by Use of an Optical Fiber',
Phys. Rev. Lett. \textbf{57}, 937--940 (1986).
\bibitem{panch_1956} S. Pancharatnam, `Generalized Theory of
Interference, and its Applications', Proc. Ind. Acad.
Sciences \textbf{A44}, 247--262 (1956).
\bibitem{epr_1935} A. Einstein, B. Podolsky and N. Rosen, `Can
quantum-mechanical description of reality be considered
complete?', Phys. Rev. \textbf{47}, 777--780 (1935).
\bibitem{schro_1936} E. Schr\"{o}dinger, `Probability relations between
separated systems', Proc. Camb. Phil. Soc. \textbf{32},
446--452 (1936).
\bibitem{neilsen_2000} For an excellent presentation of these ideas,
see, for instance, M. A. Nielsen and I. L. Chuang, `Quantum
Computation and Quantum Information', Cambridge University
Press (2000).
\bibitem{simon_2010} B. Neethi Simon, Sudhavathani Simon, F. Gori,
Massimo Santarsiero, Riccardo Borghi, N. Mukunda and R.
Simon, `Non-quantum Entanglement Resolves a Basic Issue in
Polarization Optics', Phys. Rev. Lett. \textbf{104}, 023901
-- 1 to 4 (2010); `A Complete Characterization of
Pre-Mueller and Mueller Matrices in Polarization Optics', J.
Opt. Soc. Am. \textbf{A27}, 188–-199 (2010).
\bibitem{xiao_2011}  See for instance, Q. Xiao-Feng and J. H.
Eberly, `Entanglement and Classical Polarization States',
Opt. Lett., \textbf{36}, 4110--4112 (2011); P. Ghose and A.
Mukherjee, `Entanglement in Classical Optics', Rev. Theor.
Sci., \textbf{2}, 1--14 (2014); Andrea Aiello, Falk
T\"{o}ppel, Christoph Marquardt, Elisabeth Giacobino and
Gerd Leuchs, `Classical entanglement: Oxymoron or resource?'
arXiv: 1409.0123v2 [quant-phj4], Dec. 2014.
\bibitem{seigman_1986} See for instance, A. E. Siegman, `Lasers', Univ.
Science Books (1986), Chapters 16, 17.
\bibitem{simon_1984} See for instance, R. Simon, E. C. G. Sudarshan
and N. Mukunda, `Generalized rays in first-order optics:
Transformation properties of Gaussian Schell-model fields',
Phys. Rev. \textbf{A29}, 3273--3279 (1984).
\bibitem{simon_1985} These beams were first introduced and studied
in R. Simon, E. C. G. Sudarshan and N. Mukunda, `Anisotropic
Gaussian Schell model beams: Passage through optical systems
and associated invariants', Phys. Rev. \textbf{A31},
2419--2434 (1985).
\bibitem{simon_1987} See Ref. (11) above. See also R. Simon, E. C.
G. Sudarshan and N. Mukunda, `Gaussian Wigner distributions
in quantum mechanics and optics', Phys. Rev. \textbf{A36},
3868--3880 (1987); `Gaussian Wigner distributions: a
complete characterization', Phys. Lett., \textbf{A 124},
223--228 (1982); Arvind, B. Dutta, N. Mukunda and R. Simon,
`The Real Symplectic Groups in Quantum Mechanics and
Optics', Pramana -- Journal of Physics \textbf{45}, 471--497
(1995). For a comprehensive review, see R. Simon and N.
Mukunda, `Optical phase space, Wigner representation, and
invariant quality parameters', J. Opt. Soc. Am.
\textbf{A17}, 2440--2463 (2000).
\bibitem{simon_1993} These beams were first introduced and studied
in R. Simon and N. Mukunda, `Twisted Gaussian Schell--Model
Beams', J. Opt. Soc. Am., \textbf{A10}, 95--109 (1993). For
their experimental realisation, see A. T. Friberg, E.
Tervonen and J. Turunen, `Interpretation and experimental
demonstration of twisted Gaussian Schell model beams', J.
Opt. Soc. Am. \textbf{A11}, 1818--1826 (1994).
\bibitem{simon_2000} R. Simon, `Peres--Horodecki separability
criterion for continuous variable systems', Phys. Rev.
Lett., \textbf{84}, 2726--2739 (2000).
\bibitem{gori_1994} D. Ambrosini, V. Bagini, F. Gori and M.
Santarsiero, `Twisted Gaussian Schell-model beams: a
superposition model', Jour. of Mod. Opt. \textbf{41},
1391--1399 (1994).
\end{thebibliography}
\end{document}